# Grid-level impacts of renewable energy on thermal generation: efficiency, emissions and flexibility

Dhruv Suri[1], Jacques de Chalendar[1] and Ines Azevedo[1]

[1]Department of Energy Science & Engineering, Stanford University

Wind and solar generation constitute an increasing share of electricity supply globally. We find that this leas to shifts in the operational dynamics of thermal power plants. Using fixed-effects panel regression across seven major U.S. balancing authorities, we analyze the impact of renewable generation on coal, natural gas combined cycle plants, and natural gas combustion turbines. Wind generation consistently displaces thermal output, while solar's effects vary significantly by region, achieving substantial displacement in areas with high solar penetration such as the California Independent System Operator but limited impacts in coal-reliant grids such as the Midcontinent Independent System Operator. Renewable energy sources effectively reduce carbon dioxide emissions in regions with flexible thermal plants, achieving displacement effectiveness as high as 102% in the California Independent System Operator and the Electric Reliability Council of Texas. However, in coal-heavy areas such as the Midcontinent Independent System Operator and the Pennsylvania-New Jersey-Maryland Interconnection, inefficiencies from ramping and cycling reduce carbon dioxide displacement to as low as 17% and often lead to elevated nitrogen oxides and sulfur dioxide emissions. For instance, in the Pennsylvania-New Jersey-Maryland Interconnection, sulfur dioxide displacement effectiveness drops below 50%, reflecting the challenges of integrating renewables without significant upgrades to the thermal fleet. These findings underscore the critical role of grid design, fuel mix, and operational flexibility in shaping the emissions benefits of renewables. Targeted interventions, including retrofitting high-emitting plants and deploying energy storage, are essential to maximize emissions reductions and support the decarbonization of electricity systems.

## 1. Introduction

As wind and solar generation continue to expand their footprint in the United States, these renewable sources alter the operational characteristics of legacy thermal power systems, leading to variable effects on emissions intensity, generation efficiency, and overall system stability [1]. Understanding these operational shifts is critical, not only for optimizing the efficiency of fossil-based energy generation as it adapts to support RES but also for informing future policy and renewable portfolio standards that seek to balance economic, environmental, and health objectives.

The inherent variability and intermittency of RES generation requires thermal units to operate in part-load conditions, deviating from their designed optimal performance levels [2]. This operational paradigm shift has profound implications for emissions intensity, particularly concerning carbon dioxide ($CO_2$), sulfur dioxide ($SO_2$), and nitrogen oxides ($NO_x$), which are precursors to fine particulate matter ($PM_{2.5}$) and ground-level ozone formation—pollutants with established adverse health effects [3]. Part-load operation of thermal power plants, especially coal-fired units, is associated with decreased combustion efficiency, leading to elevated specific emissions of $CO_2$, $SO_2$, and $NO_x$ per unit of electricity generated [4, 5]. Empirical studies have demonstrated that coal-fired power plants exhibit 3.4-10% higher emissions intensity during part-load operations compared to full-load conditions [6, 7, 8, 9]. This increase is attributed to suboptimal fuel combustion and the diminished efficacy of pollution control systems under reduced load scenarios [10, 11]. Furthermore, the frequent cycling and ramping necessitated by renewable energy variability not only exacerbate wear and tear on equipment but also significantly increase maintenance costs and emissions during transient states. Frequent cycling and

*Corresponding author(s): surid@stanford.edu*

ramping, driven by renewable energy variability, impose significant operational and environmental penalties on coal-fired power plants. Studies estimate that cycling increases maintenance costs by 2–5 percent of total operational expenses, shortens the lifespan of key components by up to 50 percent, and elevates emissions intensity by as much as 20 percent during transient operations [12, 13]. Thermodynamic analyses show efficiency losses of up to 15 percentage points during deep cycling, while ramping further amplifies fuel consumption and emissions rates, particularly for coal plants in renewable-heavy grids [14, 15]. These inefficiencies underscore the challenges of integrating renewables into grids reliant on inflexible thermal generation.

Elevated emissions of $SO_2$ and $NO_x$ are critical in the atmospheric formation of $PM_{2.5}$ and ozone. $SO_2$ undergoes oxidation to form sulfate aerosols, a significant component of $PM_{2.5}$, while $NO_x$ participates in photochemical reactions leading to both nitrate aerosols and ozone formation [16]. Increased ambient concentrations of $PM_{2.5}$ and ozone have been robustly linked to adverse health outcomes [17, 18, 19], including respiratory and cardiovascular diseases [20], and elevated mortality rates [3].

The second-order inefficiencies introduced by RES have profound implications for grid stability, economic efficiency, and decarbonization efforts. As thermal power plants increasingly operate at suboptimal loads to accommodate renewable variability, system operators face heightened challenges in maintaining grid reliability. This operational rigidity often results in greater renewable energy curtailment [21, 22], which not only undermines environmental benefits but also increases the cost of renewable integration by spreading fixed investment costs over a smaller effective output [23]. Empirical analyses reveal that curtailment rates rise sharply in systems with high renewable penetration, particularly when investments in flexibility-enhancing infrastructure such as storage or advanced forecasting tools are lacking [24].

These inefficiencies extend beyond immediate operational impacts to affect the broader economic landscape of renewable transitions. Elevated curtailment inflates the levelized cost of electricity (LCOE) for renewables, while the diminished efficiency of thermal power plants raises system-wide operational costs [25, 26]. Consequently, achieving renewable energy targets requires higher financial outlays than initially anticipated, which risks stalling progress on decarbonization pathways. However, till date, no study has looked longitudinally and empirically at historical generation and emissions from thermal power plants, and attributed variability in emissions and emissions intensities to renewable generation-dependent elasticities.

In this study, we analyze the impact of RES on emissions and efficiency across seven major U.S. balancing authorities (BAs), using fixed-effects panel regression models applied to granular, plant-level operational data. These balancing regions collectively accounted for two-thirds of U.S. electricity load in 2023. Specifically, we quantify how the integration of RES affects the generation, emissions, and emissions intensity of thermal power plants, with a focus on disaggregated outcomes by plant fuel type and utilization level. By examining the plant-specific responses to wind and solar generation, we aim to provide empirical insights into the resulting emissions penalties or gains that arise from suboptimal operation in low-utilization scenarios—a nuance often overlooked in broader system-level studies. Furthermore, leveraging detailed plant-level elasticities, we identify the most inefficient plants that disproportionately contribute to emissions penalties under high RES penetration. These plants, typically mid-utilization coal or older gas units, are prime candidates for targeted retrofits such as direct air capture (DAC) systems or advanced efficiency upgrades. Implementing such interventions would reduce the unit cost of abatement while improving the emissions displacement effectiveness of renewables. Our findings shed light on the regionally heterogeneous impacts of RES on fossil generation, underscoring the need for BA-specific strategies that consider fuel mix and operational dynamics unique to each area. Additionally, by establishing a framework to assess emissions scenarios under varying renewable penetrations, this study supports policymakers in evaluating the nuanced implications of renewable expansion on emissions and system efficiency across the U.S. power sector.

## 2. Results

### 2.1. Alternate $CO_2$ emissions scenarios for thermal power plants in the US

The increasing penetration of intermittent wind and solar generation imposes operational constraints on thermal power plants, leading to variable emissions outcomes. To quantify the impact of part-load inefficiencies on system-wide emissions, we develop two empirical scenarios: (1) a high-emissions scenario, where thermal plants operate consistently at the 90th percentile of their hourly $CO_2$ emissions intensity, and (2) a low-emissions scenario, where they operate at the 10th percentile. Plants are categorized by fuel type (coal, natural gas, etc.) and utilization levels (low, medium, and high capacity factors) to capture operational heterogeneity.

In the high-emissions scenario, system-wide emissions increase by 8.7–24.8% across BAs, driven by inefficiencies in coal and mid-utilization gas plants. Conversely, the low-emissions scenario achieves a reduction of 5.2–9.8%, highlighting the potential for targeted operational improvements. Natural gas-dominant BAs such as CAISO, ISO-NE, and NYISO exhibit significant flexibility, achieving emissions reductions of up to 14.6% in the low-emissions case. In CAISO, plants operating at capacity factors below 0.3 emitted 25.7 Mtons of $CO_2$, despite producing only 38.4% of the generation output of mid-utilization plants. ISO-NE and NYISO show similar patterns, where emissions reductions in the low-emissions scenario are driven by the efficiency of natural gas combined-cycle plants.

In coal-heavy regions such as ERCOT and MISO, part-load inefficiencies in mid-utilization coal plants contribute disproportionately to emissions penalties. In ERCOT, mid-utilization coal plants add 3.1 Mtons of $CO_2$ in the high-emissions scenario, compared to 1.4 Mtons for high-utilization plants. MISO follows a similar trend, with an overall emissions increase of 8.7% in the high-emissions case but a 7.7% decrease under low-emissions conditions. PJM, with its diverse generation mix, shows the highest variability: natural gas plants contribute a 14.9 Mton emissions penalty in the high-emissions case but yield a 9.4 Mton reduction under the low-emissions scenario. SWPP, balancing coal and natural gas, sees a 13.8% increase in emissions under the high-emissions scenario and an 8.1% decrease under the low-emissions scenario, with mid-utilization coal plants driving much of the variability.

These results highlight the critical role of operational efficiency in shaping emissions outcomes within net-zero energy systems. The significant emissions penalties associated with mid-utilization coal plants and the variability observed across BAs underscore the importance of prioritizing the phase-out or retrofitting of inefficient thermal units. BAs with flexible natural gas fleets, such as CAISO and NYISO, demonstrate the potential to significantly reduce emissions under optimized operation, while coal-reliant regions like MISO and SWPP illustrate the challenges of integrating renewables without exacerbating part-load inefficiencies. Achieving net-zero emissions will require a dual focus on improving operational efficiency for existing thermal plants and accelerating investments in grid technologies that minimize reliance on mid-utilization fossil units, which remain a key source of emissions variability. These targeted measures, grounded in the operational realities of individual BAs, are essential for aligning renewable integration with decarbonization goals.

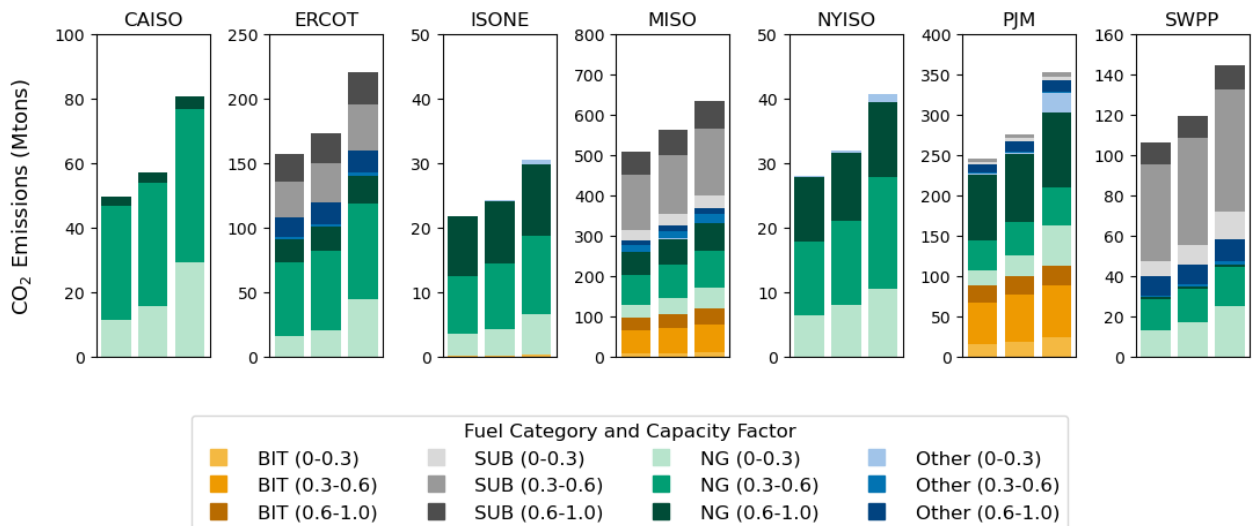

Figure 1 | **Thermal power plant $CO_2$ emissions scenarios by ISO, 2023**. Bars represent $CO_2$ emissions for thermal power plants under three scenarios: (1) a 'low emissions' scenario (10th percentile emissions intensity, left bar), (2) observed historical emissions (2023, middle bar), and (3) a 'high emissions' scenario (90th percentile emissions intensity, right bar). Y-axis limits vary by ISO to reflect regional differences. Bars are disaggregated by plant technology and annual capacity factor range, illustrating the emissions variability driven by operational and technological factors.

## 2.2. Shifting emissions profiles and generation behavior with marginal changes in renewables

Figure 2 highlights the shifting operational and emissions profiles of thermal power plants from 2010 to 2022 across BAs, revealing critical changes driven by increasing renewable generation. Panels (a) and (b) present the share of annual $CO_2$, $SO_2$, and $NO_x$ emissions relative to generation, with marker size denoting plant capacity and shading representing emissions intensity. In 2010, regions like CAISO and NYISO—dominated by natural gas—exhibited a strong positive correlation between generation share and $CO_2$ emissions, consistent with expectations. By 2022, this relationship has shifted, with larger plants showing reduced emissions intensity. For instance, in CAISO, plants contributing 0.05–0.075% of annual generation now account for a smaller proportion of $CO_2$ emissions, reflecting efficiency improvements. Similarly, in NYISO, plants generating less than 0.05% of annual output exhibit lower emissions shares, indicating the adoption of cleaner technologies and improved operational flexibility.

In coal-intensive regions like MISO and PJM, the trends differ markedly. For $SO_2$ and $NO_x$, emissions remain concentrated in mid-utilization coal plants, often within the 0.05–0.15% generation share range. In MISO, some coal plants generating below 0.1% of annual output contribute over 0.2% of $SO_2$ emissions in 2022, underscoring persistently high emissions intensity. PJM reflects similar dynamics, where mid-utilization plants (capacity factor 0.3–0.6) continue to exhibit inefficient sulfur and nitrogen emissions profiles. Despite marginal improvements in $CO_2$ intensity, these findings highlight a critical need for targeted emissions reduction strategies, particularly for coal-dominant plants that remain resistant to operational efficiency gains.

Panel (c) of Figure 2 illustrates daily marginal changes in wind and solar generation across BAs, expressed as percentage shifts from the previous day's renewable output. CAISO and ERCOT, leading in renewable integration, display highly variable and skewed distributions, with frequent large positive and negative shifts. Solar generation in these regions often experiences substantial daily increases, particularly following lower-output periods, evidenced by peaks in the 50–100% range. Wind generation, while less extreme, exhibits similar variability, reflecting the combined effects of resource intermittency and growing renewable capacity. These dynamics emphasize the increasing demands on thermal plants for flexibility, as they must adapt to rapid and unpredictable changes in renewable output.

These observed shifts in emissions and operational behavior underscore the need for grid-level strategies to manage the growing penetration of renewables. The subsequent panel regression analysis quantifies the impact of these marginal renewable changes on thermal plant operations, emissions, and efficiency, offering actionable insights for aligning thermal generation strategies with the evolving demands of a renewable-heavy grid.

## 2.3. Marginal impacts of wind and solar generation on thermal power plant production and emissions

To evaluate how increasing renewable energy affects thermal power plant operations and emissions, we employ a fixed-effects panel regression model with a logarithmic specification. This approach not only quantifies the percent change in thermal plant generation, emissions, and emissions intensity per 1% increase in wind or solar generation but also sheds light on the systemic inefficiencies and challenges associated with renewable integration. These results reveal significant heterogeneity across BAs, driven by variations in fuel mix, renewable penetration, and operational flexibility. Importantly, they underscore that while renewables reduce total thermal generation and emissions, their effectiveness is far from uniform. Without targeted interventions, these operational inefficiencies may constrain the emissions benefits of renewable energy and hinder progress toward decarbonization.

In Figure 3, we present the coefficients of solar and wind generation from the fixed-effects panel regression formulation, with generation, $CO_2$, $SO_2$, and $NO_x$ emissions intensity as dependent variables. These coefficients quantify the marginal impacts of wind and solar on thermal plant operations and emissions across BAs. Detailed regression outputs, including standard errors, confidence levels, and R-squared values for generation, $CO_2$ emissions, and $CO_2$ emissions intensity, are provided in Table 1. The remaining coefficients, capturing impacts on $SO_2$ and $NO_x$ emissions and their respective intensities, are included in the SI for comprehensive reference.

Wind generation consistently displaces thermal generation across all BAs, with coefficients ranging from -0.215 in CAISO to -0.439 in SWPP. This indicates that for every 1% increase in wind generation, thermal output declines by 0.215% and 0.439%, respectively, highlighting wind's broad compatibility with existing thermal fleets. Solar, in contrast, shows more variable impacts. In CAISO, where solar accounted for X% of total generation in 2023, a 1% increase in solar generation reduces thermal output by 0.262%. However, in

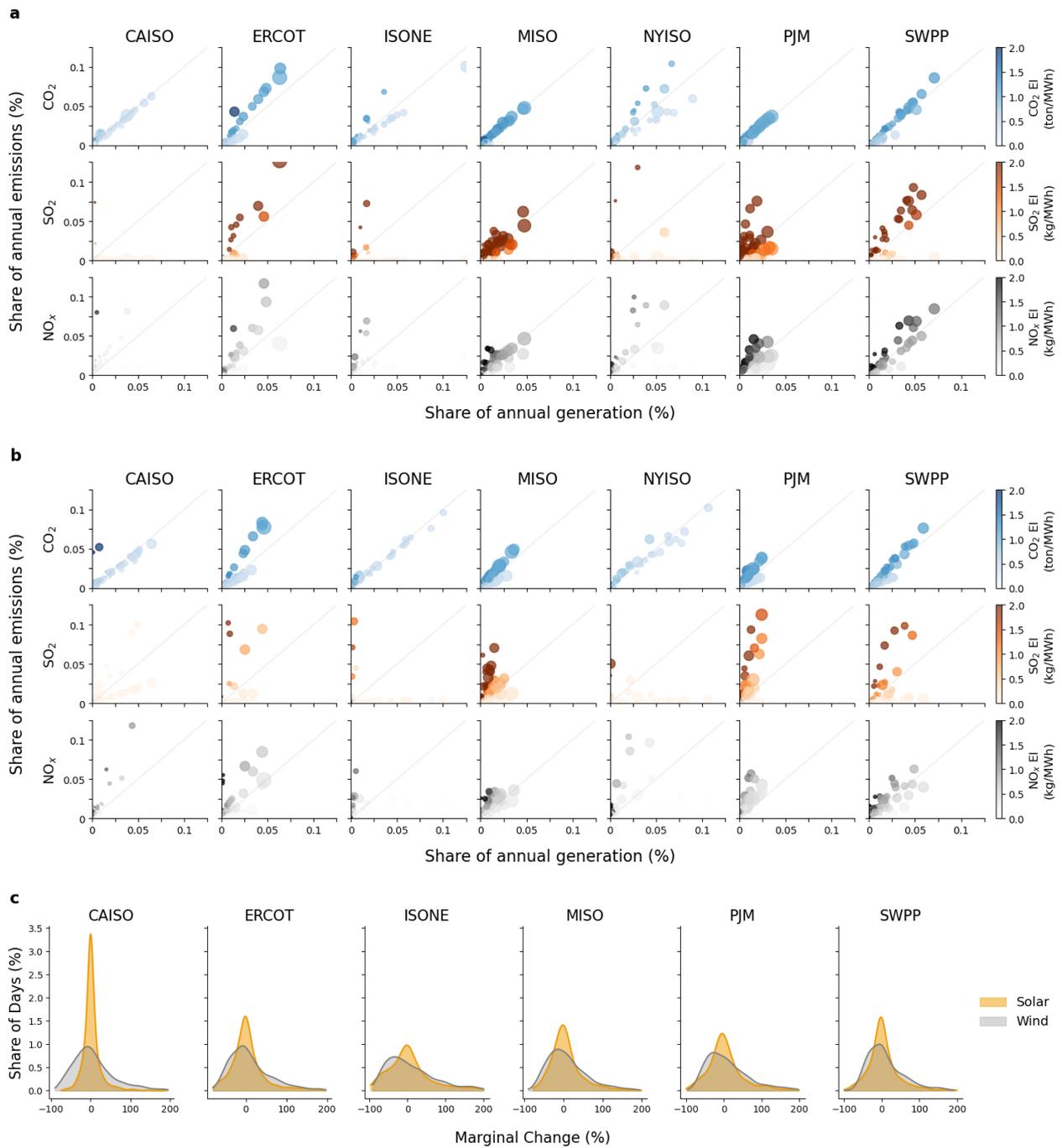

Figure 2 | **Thermal power plant emissions, generation shares, and renewable variability across seven BAs in 2010 and 2022**. **(a)** Annual $CO_2$, $SO_2$, and $NO_x$ emissions shares (y-axis) relative to annual generation shares (x-axis) for individual thermal power plants in 2010. The size of each data point corresponds to plant nameplate capacity, and the color gradient represents emissions intensity, illustrating variations across plant capacities and fuel types. **(b)** Equivalent representation for 2022, capturing shifts in emissions intensity and generation profiles over time. Natural gas-dominant regions such as CAISO and NYISO exhibit smaller and lighter data points, indicative of improved emissions efficiency. In contrast, coal-reliant BAs like MISO and PJM continue to exhibit significant $SO_2$ and $NO_x$ emissions from mid-utilization plants, underscoring coal's inherent inefficiencies at reduced loads. **(c)** Density plots of daily marginal changes in wind and solar generation, expressed as percentage shifts from the previous day's renewable output. The distributions highlight the variability in renewable generation, with CAISO and ERCOT showing pronounced daily fluctuations, particularly for solar generation.

PJM, SWPP, and ERCOT—regions with relatively lower solar penetration—the impact is negligible or even slightly positive. For example, in ERCOT, where solar contributes only X% of annual generation, thermal plants maintain higher output levels to meet load variability, limiting solar's displacement potential. These differences illustrate how renewable generation outcomes are shaped by regional renewable penetration and the operational characteristics of the thermal fleet.

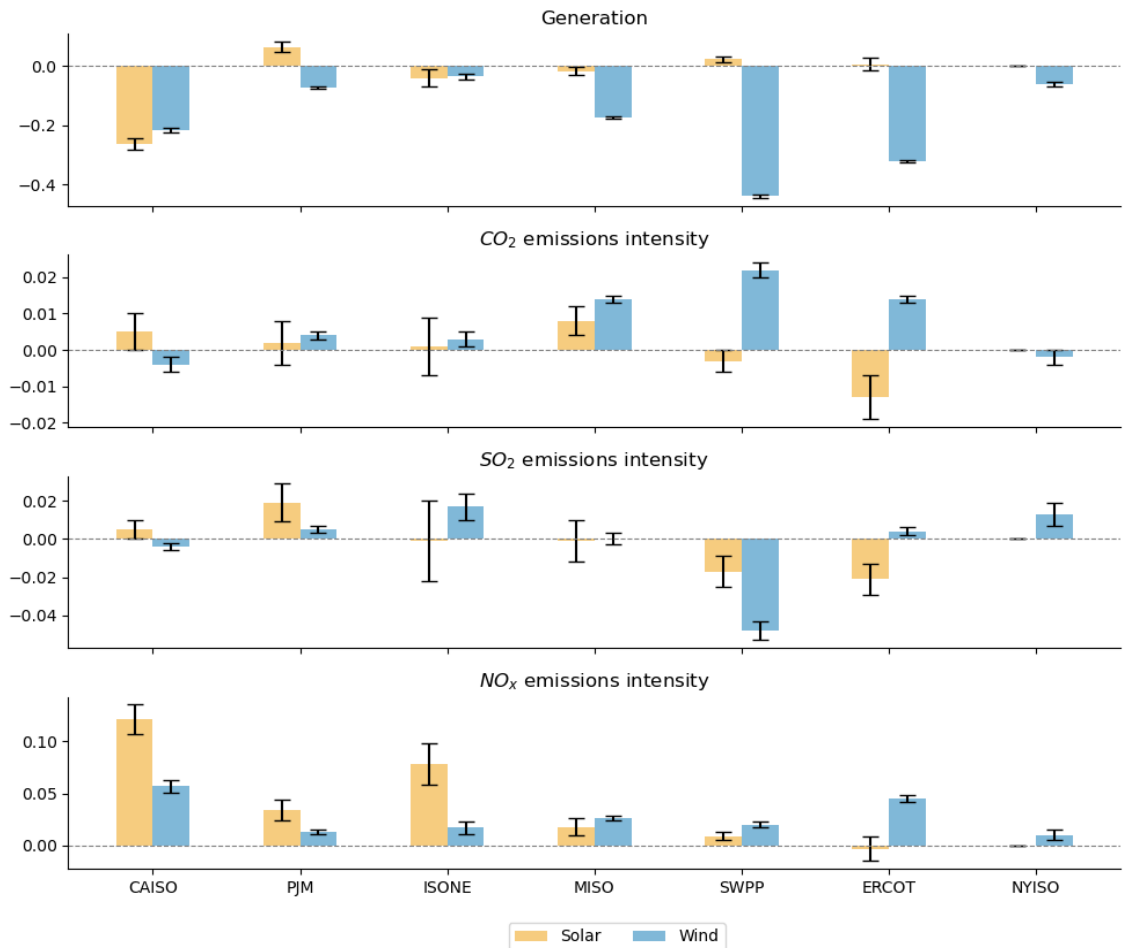

Figure 3 | **Coefficients of solar and wind generation in the fixed-effects panel regression model.** The panels illustrate the estimated effects of marginal changes in solar (yellow) and wind (blue) generation on thermal plant generation, $CO_2$, $SO_2$, and $NO_x$ emissions intensity across balancing authorities (BAs). Bars represent the point estimates of regression coefficients, quantifying the percentage change in the dependent variable per 1% increase in solar or wind generation. Error bars denote the standard error of the estimates, providing a statistical measure of uncertainty.

The emissions displacement effectiveness of renewables, summarized in Table 2, highlights this uneven performance. For $CO_2$, solar achieves high displacement effectiveness in CAISO, PJM, and SWPP (92–102%), reflecting the dominance of flexible natural gas plants, which efficiently accommodate solar variability. In contrast, ISO-NE (17%), MISO (76%), and ERCOT (33%) exhibit markedly lower effectiveness. This is due to the prevalence of coal and older gas plants, which struggle to adapt to solar generation, leading to increased emissions during cycling and start-up phases. Displacement effectiveness for wind and solar can exceed 100%. This occurs due to system-wide secondary effects, such as improved operational efficiency and reduced ramping of higher-intensity plants. A detailed mathematical explanation for this phenomenon is provided in the SI.

Wind generation demonstrates more consistent displacement effectiveness for $CO_2$, ranging from 90–103% across all BAs. Regions like CAISO and PJM benefit from flexible combined-cycle gas turbines, which are better suited to balance wind variability, whereas coal-reliant grids such as MISO and SWPP face challenges due to the inefficiencies of cycling coal plants. These inefficiencies are particularly evident in the displacement of $SO_2$ and $NO_x$, where solar and wind show reduced effectiveness in coal-heavy BAs. For example, in MISO,

Table 1 | **Coefficients of the panel regression formulation with generation, $CO_2$ emissions, and $CO_2$ emissions intensity as the dependent variable.** Significance levels: *** $p < 0.001$, ** $p < 0.01$, * $p < 0.05$.

(a) Generation

|  | **CAISO** | **PJM** | **ISONE** | **MISO** | **SWPP** | **ERCOT** | **NYISO** |
|---|---|---|---|---|---|---|---|
| **Thermal Generation** | 1.597*** | 0.171*** | 1.297*** | 1.365*** | 1.476*** | 1.864*** | 1.469*** |
|  | (0.027) | (0.009) | (0.036) | (0.022) | (0.033) | (0.028) | (0.029) |
| **Solar** | −0.262*** | 0.064*** | −0.040 | −0.018 | 0.023** | 0.006 | −0.000*** |
|  | (0.020) | (0.018) | (0.029) | (0.014) | (0.009) | (0.021) | (0.000) |
| **Wind** | −0.215*** | −0.072*** | −0.035*** | −0.175*** | −0.439*** | −0.321*** | −0.061*** |
|  | (0.008) | (0.004) | (0.009) | (0.004) | (0.006) | (0.005) | (0.008) |
| **Solar Ramp** | 0.129*** | −0.058** | 0.024 | 0.011 | −0.065*** | −0.110*** | 0.000 |
|  | (0.023) | (0.021) | (0.031) | (0.017) | (0.011) | (0.026) | (0.000) |
| **Wind Ramp** | 0.124*** | 0.040*** | −0.020 | 0.073*** | 0.108*** | 0.109*** | 0.061*** |
|  | (0.011) | (0.006) | (0.015) | (0.005) | (0.007) | (0.008) | (0.013) |
| **R-Squared** | 0.80 | 0.70 | 0.71 | 0.74 | 0.69 | 0.76 | 0.78 |
| **Num Of Obs** | 73,833 | 205,397 | 43,937 | 213,241 | 116,296 | 123,576 | 62,972 |

(b) $CO_2$ Emissions

|  | **CAISO** | **PJM** | **ISONE** | **MISO** | **SWPP** | **ERCOT** | **NYISO** |
|---|---|---|---|---|---|---|---|
| **Thermal Generation** | 1.706*** | 0.194*** | 1.644*** | 1.473*** | 1.490*** | 2.074*** | 1.750*** |
|  | (0.030) | (0.009) | (0.041) | (0.025) | (0.039) | (0.032) | (0.035) |
| **Solar** | −0.262*** | 0.079*** | −0.000 | −0.024 | 0.036*** | 0.006 | 0.000 |
|  | (0.022) | (0.018) | (0.032) | (0.016) | (0.010) | (0.024) | (0.000) |
| **Wind** | −0.241*** | −0.071*** | −0.030** | −0.178*** | −0.489*** | −0.351*** | −0.078*** |
|  | (0.009) | (0.004) | (0.010) | (0.004) | (0.007) | (0.006) | (0.010) |
| **Solar Ramp** | 0.146*** | −0.069** | 0.000 | 0.020 | −0.090*** | −0.108*** | 0.000 |
|  | (0.026) | (0.022) | (0.035) | (0.020) | (0.013) | (0.030) | (0.000) |
| **Wind Ramp** | 0.152*** | 0.041*** | −0.026 | 0.076*** | 0.159*** | 0.139*** | 0.074*** |
|  | (0.012) | (0.006) | (0.016) | (0.006) | (0.008) | (0.010) | (0.015) |
| **R-Squared** | 0.76 | 0.68 | 0.62 | 0.71 | 0.63 | 0.74 | 0.69 |
| **Num Of Obs** | 74,508 | 216,616 | 43,962 | 219,252 | 113,949 | 124,406 | 63,937 |

(c) $CO_2$ Emissions Intensity

|  | **CAISO** | **PJM** | **ISONE** | **MISO** | **SWPP** | **ERCOT** | **NYISO** |
|---|---|---|---|---|---|---|---|
| **Thermal Generation** | 0.044*** | 0.003 | −0.019 | −0.036*** | −0.057*** | −0.060*** | 0.078*** |
|  | (0.007) | (0.003) | (0.010) | (0.007) | (0.011) | (0.007) | (0.008) |
| **Solar** | 0.005 | 0.002 | 0.001 | 0.008 | −0.003 | −0.013* | 0.000 |
|  | (0.005) | (0.006) | (0.008) | (0.004) | (0.003) | (0.006) | (0.000) |
| **Wind** | −0.004* | 0.004** | 0.003 | 0.014*** | 0.022*** | 0.014*** | −0.002 |
|  | (0.002) | (0.001) | (0.002) | (0.001) | (0.002) | (0.001) | (0.002) |
| **Solar Ramp** | 0.007 | −0.004 | 0.002 | −0.009 | 0.003 | 0.010 | 0.000 |
|  | (0.006) | (0.007) | (0.008) | (0.005) | (0.003) | (0.007) | (0.000) |
| **Wind Ramp** | 0.012*** | −0.001 | 0.001 | −0.012*** | −0.011*** | −0.003 | 0.004 |
|  | (0.003) | (0.002) | (0.004) | (0.002) | (0.002) | (0.002) | (0.003) |
| **R-Squared** | 0.43 | 0.65 | 0.61 | 0.75 | 0.60 | 0.76 | 0.53 |
| **Num Of Obs** | 73,827 | 196,312 | 42,858 | 206,244 | 110,956 | 121,676 | 61,922 |

Table 2 | **Displacement effectiveness for Solar and Wind across $CO_2$, $SO_2$, and $NO_x$ emissions across ISOs.**

(a) Solar Displacement Effectiveness

|         | CAISO | PJM  | ISONE | MISO | SWPP | ERCOT | NYISO |
|---------|-------|------|-------|------|------|-------|-------|
| $CO_2$  | 0.98  | 1.02 | 0.17  | 0.76 | 0.92 | 0.33  | 0.86  |
| $SO_2$  | 0.98  | 1.26 | 1.05  | 1.03 | 0.44 | 0.41  | 0.93  |
| $NO_x$  | 0.54  | 1.52 | -     | 0.21 | 1.23 | 0.68  | 0.26  |

(b) Wind Displacement Effectiveness

|         | CAISO | PJM  | ISONE | MISO | SWPP | ERCOT | NYISO |
|---------|-------|------|-------|------|------|-------|-------|
| $CO_2$  | 1.02  | 0.95 | 0.90  | 0.93 | 0.96 | 0.96  | 1.03  |
| $SO_2$  | 1.02  | 0.94 | 0.64  | 1.00 | 1.09 | 0.99  | 0.84  |
| $NO_x$  | 0.75  | 0.81 | 0.36  | 0.86 | 0.96 | 0.88  | 0.87  |

solar displaces only 44% of expected $SO_2$ emissions, while wind achieves 81–94%. In SWPP, similar trends are observed, underscoring the challenges of curbing sulfur emissions in regions where coal remains a significant part of the generation mix. Even in CAISO, where solar integration is highest, $NO_x$ displacement effectiveness is only 54%, reflecting the elevated emissions during ramping and start-up of natural gas plants. These findings highlight the operational complexities of integrating renewables into thermal-heavy grids.

Perhaps most critically, marginal increases in wind and solar generation lead to higher emissions intensity for thermal plants, even as total emissions decline. Across all BAs, the ramping and cycling required to balance renewable variability result in increased $CO_2$ and $NO_x$ emissions intensity. For instance, in CAISO, solar generation increases $NO_x$ intensity by 0.121%, while wind increases it by 0.057%. These second-order effects are particularly acute in coal-heavy regions like MISO and SWPP, where plant inflexibility amplifies emissions penalties. Such findings emphasize the need for targeted solutions, including retrofitting coal plants, deploying more flexible natural gas technologies, and scaling up storage to mitigate the cycling of thermal plants.

A detailed breakdown of the regression coefficients, along with their statistical significance, is provided in the SI, offering a comprehensive view of regional differences in plant responses. These results not only quantify the emissions reductions associated with renewables but also underscore the importance of addressing systemic inefficiencies. Without such interventions, the integration of renewables may yield diminishing returns in emissions reductions, complicating the path to decarbonization.

### 2.4. Individual plant response is heterogeneous by fuel type and nameplate capacity

The fleet-level averages presented earlier mask significant variability in the response of individual plants to renewable generation. To better capture this heterogeneity, we apply a time fixed-effects regression model to daily observations for individual plants, categorized by balancing authority (BA) and fuel type—coal, natural gas combined cycle (NGCC), and natural gas combustion turbines (NGCT). Figure 4 presents the mean and standard error of regression coefficients with generation and $CO_2$ intensity as dependent variables, providing insights into how plants across different regions and fuel types respond to marginal changes in wind and solar generation.

Across all BAs, the responses of thermal plants to renewables vary significantly by region and fuel type. In ERCOT, NGCC plants are particularly responsive to wind generation, with a wind coefficient of -0.38, reflecting the region's substantial wind capacity and grid integration capabilities. By contrast, coal plants in PJM exhibit a positive response to solar generation (coefficient of 0.31), suggesting limited displacement of coal, likely due to its role in providing baseload power. NGCT plants, which primarily function as peaking units, exhibit inconsistent responses. For instance, in PJM, NGCT plants show a minimal response to wind, with a coefficient of -0.09, indicating that their operations are largely decoupled from wind variability and instead driven by short-term demand peaks.

The impact of renewables on emissions intensity further highlights the heterogeneity of plant responses.

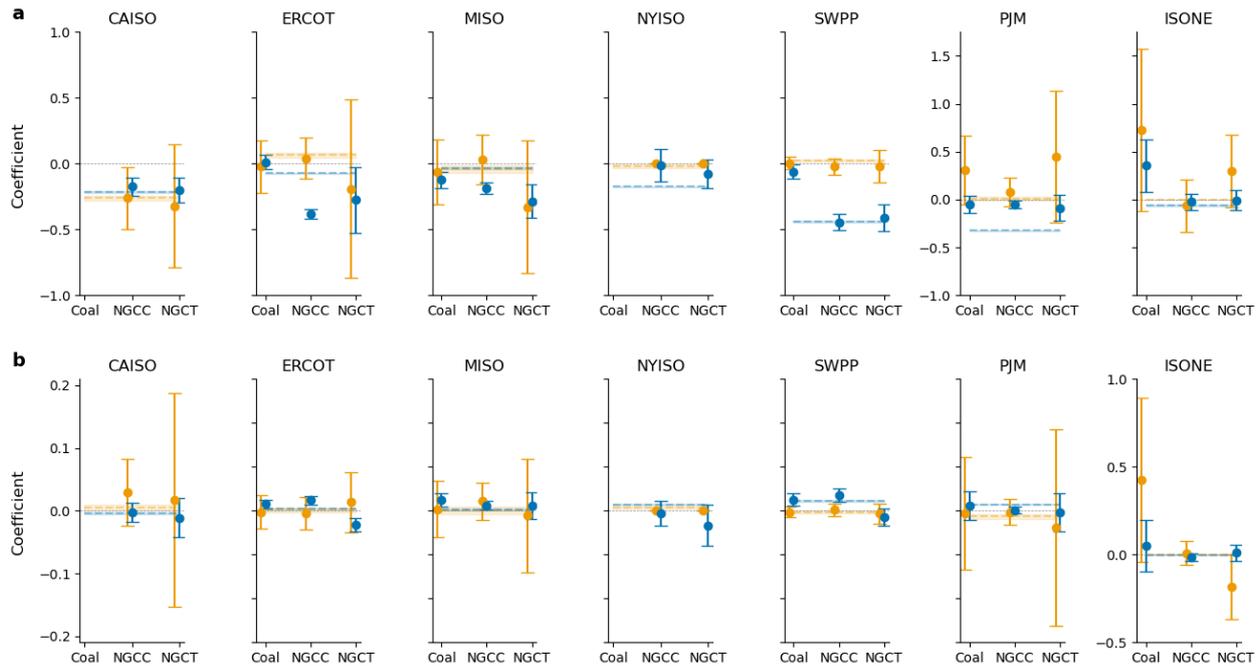

Figure 4 | **Generation and $CO_2$ Emission Intensity Coefficients for Different Fuel Types Across U.S. Electricity Markets**. (a) Coefficients representing generation response for coal, NGCC, and NGCT plants in each BA. (b) Coefficients for $CO_2$ emission intensity. The error bars represent the variance. Blue markers indicate coefficients for wind generation, while orange markers represent solar generation.

Coal plants in SWPP exhibit a notable reduction in $CO_2$ intensity with increasing wind generation, with a wind coefficient of -0.06, reflecting efficient substitution of wind for coal. Conversely, in ISONE, coal plants show an increase in $CO_2$ intensity in response to solar generation, with a coefficient of 0.43, pointing to efficiency losses from cycling and ramping. In ERCOT, coal plants show only modest decreases in emissions intensity with solar generation (coefficient of -0.006), indicating less effective displacement. NGCC plants in CAISO, known for their flexibility, exhibit near-zero emissions intensity changes, maintaining stable operations even as renewables are integrated. These contrasting outcomes highlight how fuel type, grid configuration, and operational flexibility shape the emissions response of thermal plants to renewable generation.

To translate these plant-level responses into absolute emissions reductions, we calculate the displaced tons of $CO_2$, and kilograms of $SO_2$ and $NO_x$, by multiplying plant-level coefficients by emissions during the study period and normalizing by total renewable generation. Results indicate that emissions reductions are concentrated among a small subset of plants within each BA. In ERCOT, five plants account for 21% of displaced $CO_2$, 29% of $SO_2$, and 10% of $NO_x$ emissions reductions from wind generation. In PJM, similar reductions are achieved across a broader base, with 10 plants contributing 27% of $CO_2$, 18% of $SO_2$, and 23% of $NO_x$ reductions. In coal-dominated MISO, five plants contribute 31% of displaced $CO_2$, 63% of $SO_2$, and 48% of $NO_x$. These results underscore the disproportionate role of a small number of high-emitting plants in determining regional emissions displacement.

The heterogeneity in plant responses underscores critical challenges and opportunities for grid decarbonization. In regions like ERCOT and SWPP, emissions reductions are more effective when renewable generation displaces flexible thermal plants, as evidenced by the large wind displacement coefficients for NGCC units. However, in coal-reliant grids like MISO and ISONE, inefficiencies and emissions penalties from cycling inflexible baseload plants limit the effectiveness of renewables. The concentration of emissions displacement among a small subset of plants suggests that targeted interventions—such as retrofitting high-emitting units, accelerating their retirement, or deploying storage to stabilize renewable output—could amplify the emissions benefits of renewable integration. These findings emphasize the importance of aligning renewable expansion strategies with the operational realities of thermal fleets to ensure that decarbonization efforts achieve their maximum potential.

# 3. Discussion

Our analysis highlights the operational and emissions challenges posed by the integration of renewable energy into thermal-heavy power systems. The findings reveal substantial regional variability in how renewables interact with existing thermal fleets, with critical implications for emissions reduction strategies and broader energy transitions worldwide. In regions such as CAISO and NYISO, characterized by a higher share of flexible natural gas combined cycle (NGCC) plants, the integration of solar and wind achieves high displacement effectiveness. For instance, NGCC plants in CAISO exhibit a 75% displacement effectiveness for $NO_x$ emissions, reflecting their ability to ramp efficiently in response to intermittent renewable generation. These regions demonstrate the potential for renewables to decouple electricity generation from emissions when supported by a flexible thermal fleet.

However, in coal-dependent regions like MISO and PJM, renewables face significant barriers to achieving their full emissions reduction potential. The inherent inflexibility of coal plants necessitates frequent cycling and part-load operation, leading to efficiency losses and elevated emissions intensity for $SO_2$ and $NO_x$. For example, in MISO, mid-utilization coal plants exhibit $SO_2$ displacement effectiveness of only 44–64%, far below the expected emissions reductions under a 1:1 displacement scenario. This inefficiency underscores the systemic challenges of integrating renewables into grids with substantial coal capacity, where operational dynamics can partially negate the environmental benefits of renewable energy.

Globally, these findings have broader implications for economies transitioning to renewable energy. The challenges observed in U.S. regions mirror those faced in other coal-heavy power systems, such as China, India, and parts of Europe. In these contexts, the limited flexibility of coal plants poses a major obstacle to integrating variable renewable resources. Moreover, the reliance on coal for baseload power amplifies emissions penalties during cycling events, potentially undermining national and international decarbonization targets. Addressing these issues requires a multifaceted approach, including investments in grid flexibility, targeted retrofitting of coal plants with emissions control technologies, and accelerating the deployment of flexible generation technologies like NGCC plants and energy storage.

The systematic differences in wind and solar displacement effectiveness also highlight the importance of resource-specific strategies. Wind generation consistently achieves higher displacement effectiveness due to its alignment with evening and nighttime demand patterns, complementing thermal generation's role in meeting residual load. In contrast, solar integration faces steeper challenges in coal-dominant grids, where its diurnal variability clashes with the inflexibility of coal plants. For instance, while solar achieves high $CO_2$ displacement effectiveness in CAISO and PJM (92–102%), its impact is markedly reduced in MISO and ISO-NE, where inefficiencies in coal cycling undermine emissions reductions. Globally, this suggests that solar expansion must be accompanied by substantial investments in grid storage and demand-side management to mitigate the operational penalties observed in coal-heavy systems.

The concentration of emissions reductions among a small subset of plants further underscores the need for targeted policy interventions. In ERCOT, five plants account for 21% of displaced $CO_2$ and 29% of $SO_2$ emissions from wind, while in MISO, five coal plants contribute 63% of $SO_2$ reductions. These results suggest that prioritizing retrofitting or retiring high-emitting plants could amplify the environmental benefits of renewables, especially in regions with high coal reliance. Moreover, enhancing the peaking capabilities of NGCT plants and expanding energy storage capacity would further improve system flexibility, enabling a smoother transition to high-renewable grids.

The findings of this study emphasize that renewable energy integration is not a one-size-fits-all solution. The effectiveness of renewables in reducing emissions depends heavily on the underlying grid configuration, fuel mix, and thermal plant flexibility. For global economies aiming to decarbonize their electricity sectors, the U.S. experience offers valuable lessons. Flexible grid infrastructure, regional coordination, and targeted investments in thermal fleet upgrades are essential for overcoming the inefficiencies identified in this study. As renewable penetration grows worldwide, addressing the interplay between intermittent resources and thermal power systems will be critical to achieving deep decarbonization while maintaining grid reliability and economic viability. This study provides a critical framework for understanding the operational and emissions impacts of renewable integration, extending beyond the U.S. context to inform global energy transitions. Future research should explore how these dynamics evolve with higher renewable penetrations and assess the role of emerging technologies, such as advanced energy storage and grid-interactive demand, in mitigating the challenges identified here.

# Methods

We perform our analysis for years 2018 through 2023 for seven balancing authority regions in the US. We start by describing the data used, followed by the methods.

**Data**

Our study focuses on seven major Independent System Operator (ISO) regions in the United States: the California Independent System Operator (CAISO), the Electric Reliability Council of Texas (ERCOT), the New England Independent System Operator (ISO-NE), the Midcontinent Independent System Operator (MISO), the New York Independent System Operator (NY-ISO), the Pennsylvania-New Jersey-Maryland Interconnection (PJM), and the Southwest Power Pool (SWPP). Specifically, we analyze the operational behavior of thermal power plants within these regions. The analysis uses hourly data on electricity generation and emissions, including carbon dioxide ($CO_2$), sulfur dioxide ($SO_2$), and nitrogen oxides ($NO_x$), sourced from the Environmental Protection Agency's Continuous Emissions Monitoring System (CEMS) [27]. This dataset covers all generating units with a nameplate capacity exceeding 25 megawatts.

The study spans the period from July 1, 2018, to December 31, 2023, encompassing all seven ISO regions. Hourly wind and solar power production, electricity imports, and load demand data for each ISO are retrieved from the U.S. Energy Information Administration's hourly electric grid monitor. This data has been processed using a physics-informed reconciliation method to ensure accuracy and consistency [28].

Thermal power plants included in the analysis are those with at least 10% of hourly generation and emissions data available during the study period. The final dataset comprises 1,264 thermal power plants, of which 744 are powered by natural gas, 248 by subbituminous coal, 213 by bituminous coal, and 59 by other fuel types. Together, these plants accounted for 67% of total electricity generation and over 94% of emissions across the seven ISO regions in 2023. Additional plant characteristics, such as geographic location, primary fuel type, operational age, and stack height, are sourced from the Environmental Protection Agency's Emissions & Generation Resource Integrated Database (eGRID) [29].

**Methods**

*Statistical models*

We use a fixed-effects panel regression formulation to quantify the marginal effect of wind and solar generation on the generation, emissions, and emissions intensity of thermal plants. Similar to Bushnell and Wolfram [30], Graf et al. [31] and Fell and Johnson [11], we use an entity and time fixed-effects logarithmic specification for all plants within an ISO using daily observations as follows:

$$\ln y_{i,t} = \alpha_i + \beta_1 \ln D'_t + \beta_2 \ln S_t + \beta_3 \ln W_t + \beta_4 \ln \overline{W}_t + \gamma \ln \mathbf{X}_t + \eta_{t,m} + \eta_{t,y} + \alpha_i \times \eta_{t,m} + \epsilon_i \quad (1)$$

where $y_{i,t}$ is the dependent variable of interest, which is daily generation in MWh, emissions in tonnes or emissions intensity in tonnes/MWh for power plant $i$ at timestep $t$. $D_t$, $S_t$ and $W_t$ represent the daily demand, generation from solar, and generation from wind in MWh for the ISO to which $i$ belongs. The daily net thermal demand is the total ISO thermal demand minus generation from hydro and imports, i.e., $D'_t = D_t - H_t - I_t$, where $D_t$, $H_t$ and $I_t$ represent the daily thermal demand, generation from hydro and net imports in MWh for the ISO. $\overline{W}_t$ is a parameter that estimates the intermittency of net generation from wind at the hourly level and is calculated by taking the sum of the absolute value of hourly differences in wind output over the course of the day. Mathematically, this is expressed as:

$$\overline{W}_t = \sum_{h=0}^{23} |W_{t,h+1} - W_{t,h}| \quad (2)$$

$\mathbf{X}_t$ is a set of control variables that includes the sum of wind generation, solar generation, and demand for trading partners of the ISO to which $i$ belongs. For instance, CAISO trades with the other balancing authorities in the Northwest (NW) and Southwest (SW) NERC subregions. For every hour, these external control variables would consider the sum of wind generation, solar generation, and demand for these two regions. For ERCOT, the only trading partner is the Southwest Power Pool (SWPP). $\eta_{t,m}$ and $\eta_{t,y}$ are the time fixed-effects parameters controlling for the month and year, respectively. We interact the monthly fixed-effects variable with the indicator variable for the plant ID to account for heterogeneity in fuel prices across different technology categories, regions and plant types. $\epsilon$ represents the error term.

The model specifications employed by Graf et al. [31] and Fell and Johnson [11] may not be able to fully explain the deviation caused due to intermittency in renewable generation due to gaps in granularity and panel formulation. Graf et al. [31] do not include granular regional generation from trading partners, and also conduct the analysis on annual data and hence miss intra-day operational parameters such as resource ramping. Fell and Johnson [11] do not construct a plant-level panel regression but rather run a time fixed-effects model for the BA that does not account for location-based heterogeneity.

In our formulation, the parameters that are of interest are the coefficients for solar, $\beta_2$ and wind, $\beta_3$. These coefficients represent the percent change in the dependent variable per 1% increase in generation from solar and wind, respectively. For example, if the dependent variable $y_{i,t}$ represents the net generation from thermal plants for an ISO, then $\beta_2$ and $\beta_3$ will represent the percent change in generation for thermal plants per unit percent increase in generation from solar and wind. These coefficients thus measure the associated change in the dependent variable under the assumption that daily wind and solar production (represented by $W$ and $S$) is uncorrelated with the error term $\epsilon$ after controlling for net demand, entity, and time fixed effects. As suggested by [32], this assumption is valid for wind at the hourly level and thus can be extrapolated to daily observations. Solar output is likely to be correlated with the error term and time fixed effects parameter when considering hourly data, which is why we aggregate hourly observations to daily intervals in the statistical model above. To test for correlation between thermal power plant generation and other exogenous variables, we present the Pearson's correlation coefficient in the SI, along with the Durbin-Watson test to check for autocorrelation in the panel regression model for each entity.

*Plant-level ordinary least squares regression*

Modeling the response of individual plants outside a panel regression formulation helps us understand whether plants are individually load-following, solar-following, and wind-following, and the extent of their response. In contrast to fixed effects ISO-level formulation considering daily time steps, the plant-level model uses hourly data and is given by:

$$\ln y_t = \alpha + \beta_1 \ln D'_t + \beta_2 \ln S_t + \beta_3 \ln W_t + \beta_4 \ln \overline{W}_t + \gamma \ln \mathbf{X}_t + \eta_{t,m} + \eta_{t,y} + \epsilon \qquad (3)$$

where the variables are the same as those in the panel regression formulation above. In the plant-level OLS, $\overline{W}_t$ measures the deviation in wind generation from the previous hour.

*Deviation between observed and actual emissions*

To determine the share of emissions displaced due to renewables, we use the panel regression formulations where emissions and emissions intensity are the dependent variables similar to Graf et al. [31].

$$\ln CO_2 EI_{p,t} = \alpha_i + \beta_1 \ln D'_t + \beta_2 \ln S_t + \beta_3 \ln W_t + \beta_4 \ln \overline{W}_t + \gamma \ln \mathbf{X}_t + \eta_{t,m} + \eta_{t,y} + \alpha_i \times \eta_{t,m} + \epsilon_i \qquad (4)$$

$$\ln CO_{2,p,t} = \alpha'_i + \beta'_1 \ln D'_t + \beta'_2 {}^{\text{'}}\ln S_t + \beta'_3 \ln W_t + \beta'_4 \ln \overline{W}_t + \gamma' \ln \mathbf{X}_t + \eta'_{t,m} + \eta'_{t,y} + \alpha'_i \times \eta'_{t,m} + \epsilon'_i \qquad (5)$$

In Equation 4, the marginal effect due to increased generation from solar can be represented by:

$$\alpha_2 = \frac{\partial(\ln CO_2 EI_{p,t})}{\partial(\ln solar_t)} \qquad (6)$$

$$\alpha_2 = \frac{\partial(\ln \frac{\text{CO}_{2,p,t}}{G_{p,t}})}{\partial(\ln \text{solar}_t)} \tag{7}$$

$$\alpha_2 = \frac{\partial(\ln \text{CO}_{2,p,t} - \ln G_{p,t})}{\partial(\ln \text{solar}_t)} \tag{8}$$

$$\alpha_2 = \frac{\partial(\ln \text{CO}_{2,p,t})}{\partial(\ln \text{solar}_t)} - \frac{\partial(\ln G_{p,t})}{\partial(\ln \text{solar}_t)} \tag{9}$$

$$\alpha_2 = \alpha_2' - \frac{\partial(\ln G_{p,t})}{\partial(\ln \text{solar}_t)} \tag{10}$$

Rearranging, gives:

$$\frac{\partial(\ln G_{p,t})}{\partial(\ln \text{solar}_t)} = \alpha_2 - \alpha_2' \tag{11}$$

The final equation represents the change in thermal generation if generation from solar replaces the generation from fossil-fueled plants on a 1:1 basis. Thus, the fraction of expected emissions reductions for solar is represented by:

$$\text{Fraction of expected emissions reduction (solar)} = \frac{\alpha_2'}{\alpha_2' - \alpha_2} \tag{12}$$

Similarly, for wind, the expected emissions reductions and the corresponding fraction are given by:

$$\frac{\partial(\ln G_{p,t})}{\partial(\ln \text{wind}_t)} = \alpha_3 - \alpha_3' \tag{13}$$

$$\text{Fraction of expected emissions reduction (wind)} = \frac{\alpha_3'}{\alpha_3' - \alpha_3} \tag{14}$$

## Data Availability

Source data are provided with this paper.

## Code Availability

The code to replicate the results is available at
https://github.com/dhruvsuri17/us-emissions-impacts.

## A. Emissions scenarios

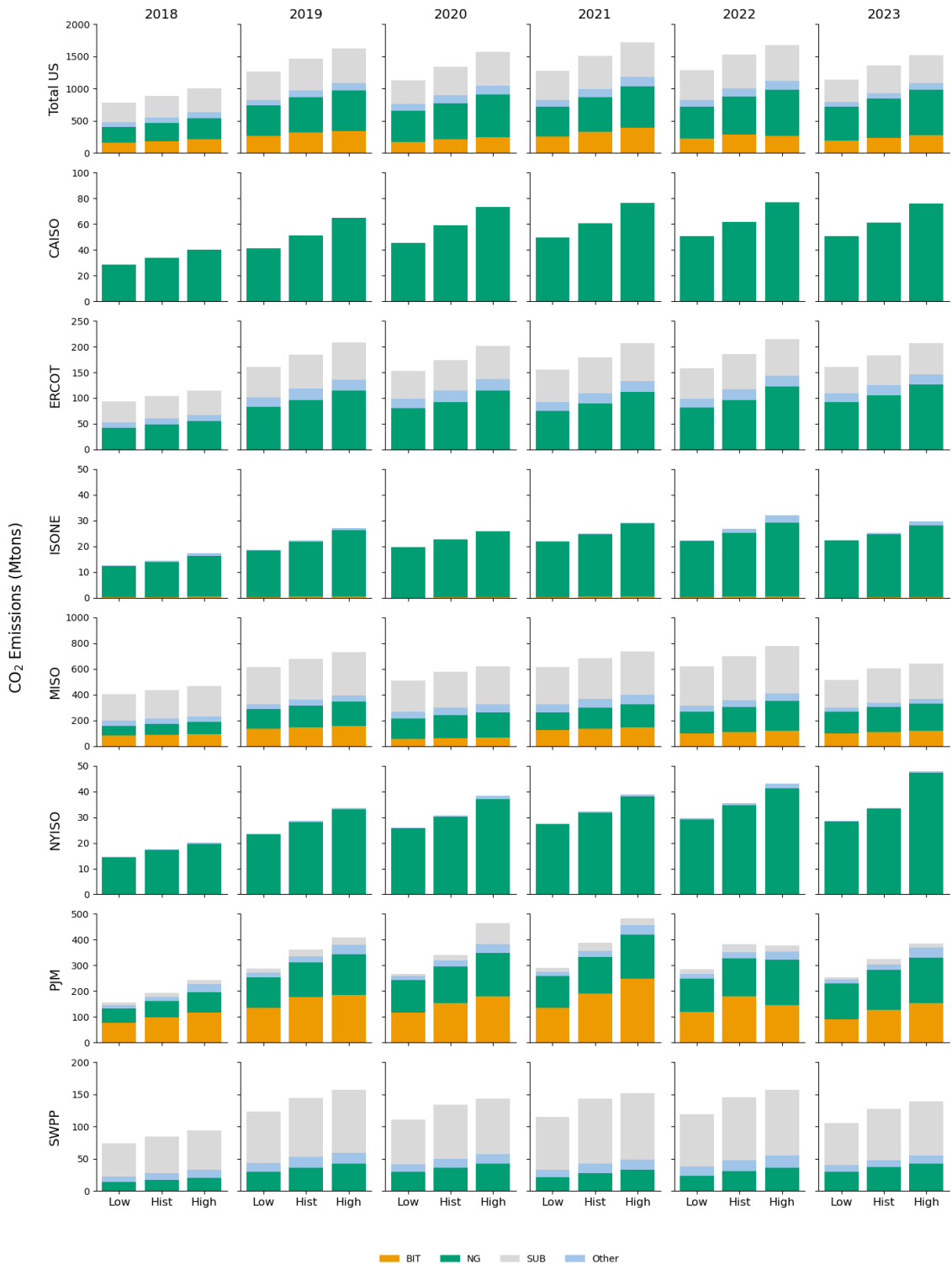

Figure 5 | **Thermal power plant emissions under two alternate emissions scenarios - US total and by ISO**. The high emissions scenario (bar on the extreme right) assumes plants operating at the 95th percentile of emissions intensity and low emissions (bar on the extreme left) assumes plants operating at the 5th percentile of emissions intensity. The middle bar denotes observed historical emissions.

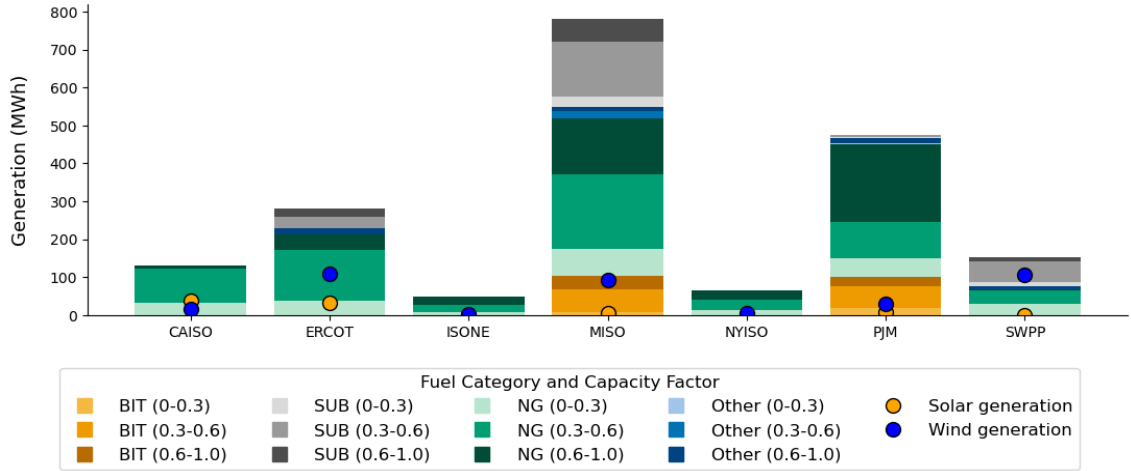

Figure 6 | **Thermal power plant emissions under two alternate emissions scenarios - US total and by ISO**. The high emissions scenario (bar on the extreme right) assumes plants operating at the 95th percentile of emissions intensity and low emissions (bar on the extreme left) assumes plants operating at the 5th percentile of emissions intensity. The middle bar denotes observed historical emissions.

## B. Regression formulation

### B.1. Treatment of zeros

The main regression specification with generation as the dependent variable is given by:

$$\ln g_{i,t} = \alpha_i + \beta_1 \ln G_t + \beta_2 \ln S_t + \beta_3 \ln W_t + \beta_4 \ln \overline{W}_t + \gamma \ln \mathbf{X}_t + \omega \ln \mathbf{Y}_t + \eta_{t,m} + \eta_{t,a} + \alpha_i \times \eta_{t,m} + \epsilon_i \quad (15)$$

Given the logarithmic nature of the regression specification, two issues arise. The first is that variables such as plant generation may be zero for a given day and it is not possible to take the logarithm of zero. Second, imports defined in $\mathbf{Y}_t$ may be positive or negative, depending on the net volume of electricity that is imported or exported by the ISO. The latter is easy to address: instead of defining net imports, we define imports, $I_t$, and exports, $E_t$, which are both positive variables. The sum of these two variables does not yield the net imports, but rather, represents whether the aggregated daily net imports were positive or negative. To address the former issue of zeros in the dependent and independent variables, we add a positive constant $\Delta$, to all observations $Y_i$ so that $\tilde{Y}_i = Y_i + \Delta > 0$. Since the choice of $\Delta$ is discretionary and may bias the estimates and their standard errors, we evaluate the coefficient estimates over a range of values for $\Delta$ and report their standard errors, model R-squared values, and p-values.

The updated model formulation is then given by:

$$\ln g_{i,t} = \alpha_i + \beta_1 \ln G_t + \beta_2 \ln S_t + \beta_3 \ln W_t + \beta_4 \ln \overline{W}_t + \gamma \ln \mathbf{X}_t + \omega_1 \ln H_t + \omega_2 \ln I_t + \\ \omega_3 \ln E_t + \eta_{t,m} + \eta_{t,a} + \alpha_i \times \eta_{t,m} + \epsilon_i$$

Adding $\Delta > 0$ to power plant generation, imports, exports and hydro in the equation above yields:

$$\ln(g_{i,t} + \Delta) = \alpha_i + \beta_1 \ln G_t + \beta_2 \ln S_t + \beta_3 \ln W_t + \beta_4 \ln \overline{W}_t + \gamma \ln \mathbf{X}_t + \omega_1 \ln(H_t + \Delta) + \omega_2 \ln(I_t + \Delta) + \\ \omega_3 \ln(E_t + \Delta) + \eta_{t,m} + \eta_{t,a} + \alpha_i \times \eta_{t,m} + \epsilon_i$$

For the dependent variable $g_{i,t}$, we use the following logarithmic identity:

Table 3 | **Coefficient of the panel regression formulation with $CO_2$, $SO_2$, and $NO_x$ emissions as the dependent variable.** Significance levels: *** $p < 0.001$, ** $p < 0.01$, * $p < 0.05$.

(a) $CO_2$ Emissions

|  | **CAISO** | **PJM** | **ISONE** | **MISO** | **SWPP** | **ERCOT** | **NYISO** |
|---|---|---|---|---|---|---|---|
| **Thermal Generation** | 1.706*** | 0.194*** | 1.644*** | 1.473*** | 1.490*** | 2.074*** | 1.750*** |
|  | (0.030) | (0.009) | (0.041) | (0.025) | (0.039) | (0.032) | (0.035) |
| **Solar** | −0.262*** | 0.079*** | −0.000 | −0.024 | 0.036*** | 0.006 | 0.000 |
|  | (0.022) | (0.018) | (0.032) | (0.016) | (0.010) | (0.024) | (0.000) |
| **Wind** | −0.241*** | −0.071*** | −0.030** | −0.178*** | −0.489*** | −0.351*** | −0.078*** |
|  | (0.009) | (0.004) | (0.010) | (0.004) | (0.007) | (0.006) | (0.010) |
| **Solar Ramp** | 0.146*** | −0.069** | 0.000 | 0.020 | −0.090*** | −0.108*** | 0.000 |
|  | (0.026) | (0.022) | (0.035) | (0.020) | (0.013) | (0.030) | (0.000) |
| **Wind Ramp** | 0.152*** | 0.041*** | −0.026 | 0.076*** | 0.159*** | 0.139*** | 0.074*** |
|  | (0.012) | (0.006) | (0.016) | (0.006) | (0.008) | (0.010) | (0.015) |
| **R-Squared** | 0.76 | 0.68 | 0.62 | 0.71 | 0.63 | 0.74 | 0.69 |
| **Num Of Obs** | 74,508 | 216,616 | 43,962 | 219,252 | 113,949 | 124,406 | 63,937 |

(b) $SO_2$ Emissions

|  | **CAISO** | **PJM** | **ISONE** | **MISO** | **SWPP** | **ERCOT** | **NYISO** |
|---|---|---|---|---|---|---|---|
| **Thermal Generation** | 1.697*** | 0.214*** | 1.702*** | 1.562*** | 1.807*** | 2.139*** | 1.971*** |
|  | (0.030) | (0.010) | (0.048) | (0.030) | (0.048) | (0.032) | (0.039) |
| **Solar** | −0.263*** | 0.090*** | −0.031 | −0.037 | 0.013 | 0.015 | 0.000 |
|  | (0.022) | (0.020) | (0.038) | (0.019) | (0.013) | (0.024) | (0.000) |
| **Wind** | −0.240*** | −0.072*** | −0.030** | −0.193*** | −0.558*** | −0.358*** | −0.065*** |
|  | (0.009) | (0.005) | (0.012) | (0.005) | (0.009) | (0.006) | (0.011) |
| **Solar Ramp** | 0.148*** | −0.071** | 0.023 | 0.035 | −0.077*** | −0.118*** | 0.000 |
|  | (0.026) | (0.024) | (0.041) | (0.024) | (0.016) | (0.031) | (0.000) |
| **Wind Ramp** | 0.150*** | 0.044*** | −0.003 | 0.087*** | 0.146*** | 0.128*** | 0.041* |
|  | (0.012) | (0.006) | (0.019) | (0.007) | (0.010) | (0.010) | (0.017) |
| **R-Squared** | 0.76 | 0.89 | 0.53 | 0.90 | 0.86 | 0.91 | 0.65 |
| **Num Of Obs** | 74,486 | 218,350 | 41,264 | 223,245 | 118,576 | 120,448 | 62,418 |

(c) $NO_x$ Emissions

|  | **CAISO** | **PJM** | **ISONE** | **MISO** | **SWPP** | **ERCOT** | **NYISO** |
|---|---|---|---|---|---|---|---|
| **Thermal Generation** | 1.167*** | 0.178*** | 1.338*** | 1.478*** | 1.598*** | 2.085*** | 1.545*** |
|  | (0.026) | (0.009) | (0.038) | (0.025) | (0.041) | (0.033) | (0.031) |
| **Solar** | −0.143*** | 0.100*** | 0.060* | −0.005 | 0.048*** | 0.007 | 0.000 |
|  | (0.019) | (0.017) | (0.031) | (0.016) | (0.011) | (0.025) | (0.000) |
| **Wind** | −0.171*** | −0.057*** | −0.009 | −0.162*** | −0.495*** | −0.330*** | −0.062*** |
|  | (0.008) | (0.004) | (0.009) | (0.004) | (0.008) | (0.006) | (0.009) |
| **Solar Ramp** | 0.099*** | −0.078*** | −0.053 | 0.003 | −0.100*** | −0.104** | 0.000 |
|  | (0.022) | (0.020) | (0.033) | (0.020) | (0.013) | (0.032) | (0.000) |
| **Wind Ramp** | 0.107*** | 0.038*** | −0.021 | 0.070*** | 0.146*** | 0.131*** | 0.060*** |
|  | (0.010) | (0.006) | (0.015) | (0.006) | (0.008) | (0.010) | (0.014) |
| **R-Squared** | 0.72 | 0.69 | 0.49 | 0.73 | 0.61 | 0.74 | 0.71 |
| **Num Of Obs** | 74,488 | 233,832 | 44,954 | 236,935 | 119,103 | 126,340 | 70,453 |

Table 4 | **Coefficient of the panel regression formulation with $CO_2$, $SO_2$, and $NO_x$ emissions intensity as the dependent variable.** Significance levels: *** $p < 0.001$, ** $p < 0.01$, * $p < 0.05$.

(a) $CO_2$ Emissions Intensity

|  | **CAISO** | **PJM** | **ISONE** | **MISO** | **SWPP** | **ERCOT** | **NYISO** |
|---|---|---|---|---|---|---|---|
| **Thermal Generation** | 0.044*** | 0.003 | −0.019 | −0.036*** | −0.057*** | −0.060*** | 0.078*** |
|  | (0.007) | (0.003) | (0.010) | (0.007) | (0.011) | (0.007) | (0.008) |
| **Solar** | 0.005 | 0.002 | 0.001 | 0.008 | −0.003 | −0.013* | 0.000 |
|  | (0.005) | (0.006) | (0.008) | (0.004) | (0.003) | (0.006) | (0.000) |
| **Wind** | −0.004* | 0.004** | 0.003 | 0.014*** | 0.022*** | 0.014*** | −0.002 |
|  | (0.002) | (0.001) | (0.002) | (0.001) | (0.002) | (0.001) | (0.002) |
| **Solar Ramp** | 0.007 | −0.004 | 0.002 | −0.009 | 0.003 | 0.010 | 0.000 |
|  | (0.006) | (0.007) | (0.008) | (0.005) | (0.003) | (0.007) | (0.000) |
| **Wind Ramp** | 0.012*** | −0.001 | 0.001 | −0.012*** | −0.011*** | −0.003 | 0.004 |
|  | (0.003) | (0.002) | (0.004) | (0.002) | (0.002) | (0.002) | (0.003) |
| **R-Squared** | 0.43 | 0.65 | 0.61 | 0.75 | 0.60 | 0.76 | 0.53 |
| **Num Of Obs** | 73,827 | 196,312 | 42,858 | 206,244 | 110,956 | 121,676 | 61,922 |

(b) $SO_2$ Emissions Intensity

|  | **CAISO** | **PJM** | **ISONE** | **MISO** | **SWPP** | **ERCOT** | **NYISO** |
|---|---|---|---|---|---|---|---|
| **Thermal Generation** | 0.043*** | 0.027*** | 0.207*** | 0.054** | 0.201*** | 0.031** | 0.361*** |
|  | (0.007) | (0.005) | (0.027) | (0.017) | (0.029) | (0.011) | (0.020) |
| **Solar** | 0.005 | 0.019 | −0.001 | −0.001 | −0.017* | −0.021* | 0.000 |
|  | (0.005) | (0.010) | (0.021) | (0.011) | (0.008) | (0.008) | (0.000) |
| **Wind** | −0.004* | 0.005 | 0.017* | −0.000 | −0.048*** | 0.004* | 0.013* |
|  | (0.002) | (0.002) | (0.007) | (0.003) | (0.005) | (0.002) | (0.006) |
| **Solar Ramp** | 0.007 | −0.017 | 0.005 | −0.007 | 0.004 | 0.012 | 0.000 |
|  | (0.006) | (0.011) | (0.023) | (0.013) | (0.009) | (0.010) | (0.000) |
| **Wind Ramp** | 0.011*** | −0.001 | 0.002 | −0.006 | −0.004 | −0.010** | −0.025** |
|  | (0.003) | (0.003) | (0.011) | (0.004) | (0.006) | (0.003) | (0.009) |
| **R-Squared** | 0.55 | 0.96 | 0.66 | 0.95 | 0.92 | 0.98 | 0.35 |
| **Num Of Obs** | 73,826 | 198,358 | 40,447 | 210,440 | 115,786 | 117,837 | 60,505 |

(c) $NO_x$ Emissions Intensity

|  | **CAISO** | **PJM** | **ISONE** | **MISO** | **SWPP** | **ERCOT** | **NYISO** |
|---|---|---|---|---|---|---|---|
| **Thermal Generation** | −0.464*** | −0.007 | −0.227*** | −0.001 | −0.021 | −0.166*** | −0.036* |
|  | (0.019) | (0.005) | (0.025) | (0.013) | (0.015) | (0.016) | (0.017) |
| **Solar** | 0.121*** | 0.034*** | 0.078*** | 0.018* | 0.009* | −0.003 | 0.000 |
|  | (0.014) | (0.010) | (0.020) | (0.008) | (0.004) | (0.012) | (0.000) |
| **Wind** | 0.057*** | 0.013*** | 0.017** | 0.026*** | 0.020*** | 0.045*** | 0.010* |
|  | (0.006) | (0.002) | (0.006) | (0.002) | (0.003) | (0.003) | (0.005) |
| **Solar Ramp** | −0.033* | −0.030** | −0.064** | −0.013 | −0.010* | 0.013 | 0.000 |
|  | (0.016) | (0.011) | (0.022) | (0.010) | (0.005) | (0.015) | (0.000) |
| **Wind Ramp** | −0.028*** | −0.002 | 0.007 | −0.019*** | −0.013*** | −0.019*** | −0.009 |
|  | (0.007) | (0.003) | (0.010) | (0.003) | (0.003) | (0.005) | (0.008) |
| **R-Squared** | 0.59 | 0.89 | 0.77 | 0.89 | 0.87 | 0.85 | 0.86 |
| **Num Of Obs** | 73,826 | 205,359 | 43,930 | 213,225 | 116,243 | 123,561 | 62,960 |

$$\ln(g_{i,t} + \Delta) = \ln(g_{i,t}) + \ln\left(1 + \frac{\Delta}{g_{i,t}}\right)$$

For the independent variables $H_t$, $I_t$, and $E_t$, adding $\Delta$ gives:

$$\ln(H_t + \Delta) = \ln(H_t) + \ln\left(1 + \frac{\Delta}{H_t}\right)$$

$$\ln(I_t + \Delta) = \ln(I_t) + \ln\left(1 + \frac{\Delta}{I_t}\right)$$

$$\ln(E_t + \Delta) = \ln(E_t) + \ln\left(1 + \frac{\Delta}{E_t}\right)$$

After adding $\Delta$ to $g_{i,t}$, $H_t$, $I_t$, and $E_t$, the model becomes:

$$\ln g_{i,t} + \ln\left(1 + \frac{\Delta}{g_{i,t}}\right) = \alpha_i + \beta_1 \ln G_t + \beta_2 \ln S_t + \beta_3 \ln W_t + \beta_4 \ln \overline{W}_t + \gamma \ln \mathbf{X}_t$$
$$+ \omega_1 \left[\ln H_t + \ln\left(1 + \frac{\Delta}{H_t}\right)\right] + \omega_2 \left[\ln I_t + \ln\left(1 + \frac{\Delta}{I_t}\right)\right] + \omega_3 \left[\ln E_t + \ln\left(1 + \frac{\Delta}{E_t}\right)\right]$$
$$+ \eta_{t,m} + \eta_{t,a} + \alpha_i \times \eta_{t,m} + \epsilon_i$$

Rearranging the model to isolate the error term:

$$\ln g_{i,t} = \alpha_i + \beta_1 \ln G_t + \beta_2 \ln S_t + \beta_3 \ln W_t + \beta_4 \ln \overline{W}_t + \gamma \ln \mathbf{X}_t$$
$$+ \omega_1 \ln H_t + \omega_2 \ln I_t + \omega_3 \ln E_t + \eta_{t,m} + \eta_{t,a} + \alpha_i \times \eta_{t,m}$$
$$+ \left[\epsilon_i + \ln\left(1 + \frac{\Delta}{g_{i,t}}\right) - \omega_1 \ln\left(1 + \frac{\Delta}{H_t}\right) - \omega_2 \ln\left(1 + \frac{\Delta}{I_t}\right) - \omega_3 \ln\left(1 + \frac{\Delta}{E_t}\right)\right]$$

Thus, the new error term is:

$$\epsilon'_i = \epsilon_i + \ln\left(1 + \frac{\Delta}{g_{i,t}}\right) - \omega_1 \ln\left(1 + \frac{\Delta}{H_t}\right) - \omega_2 \ln\left(1 + \frac{\Delta}{I_t}\right) - \omega_3 \ln\left(1 + \frac{\Delta}{E_t}\right)$$

Figure 7a shows the regression coefficients for wind and solar generation when $\Delta$ is added to $g_i$, $H_t$, $I_t$ and $E_t$. Alternatively, if we first remove daily entries where the power plant generation is 0 and add $\Delta$ for $H_t$, $I_t$ and $E_t$, the regression coefficients are shown in Figure 7b.

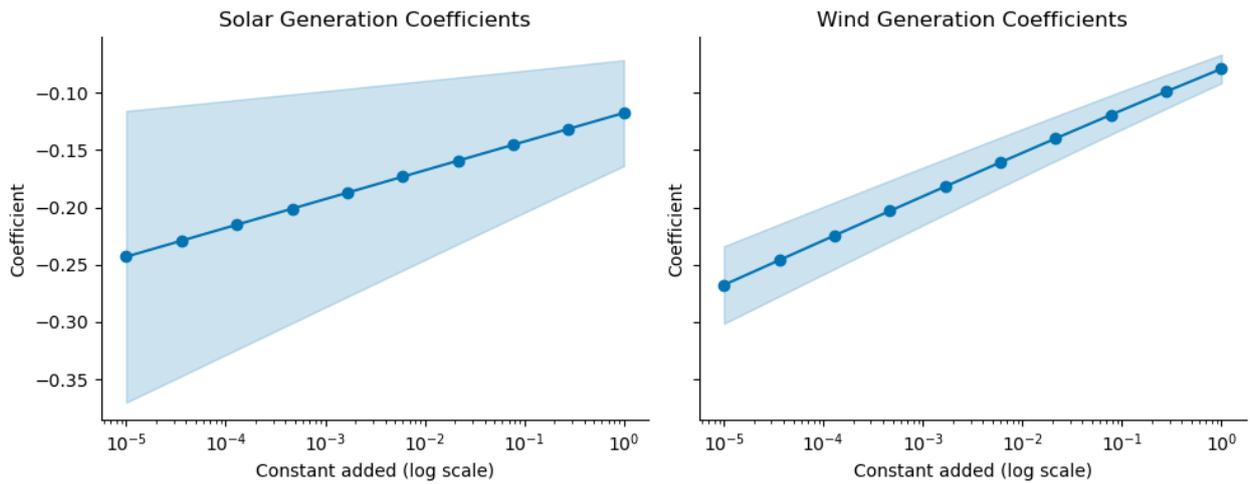

(a) All operational and non-operational hours considered.

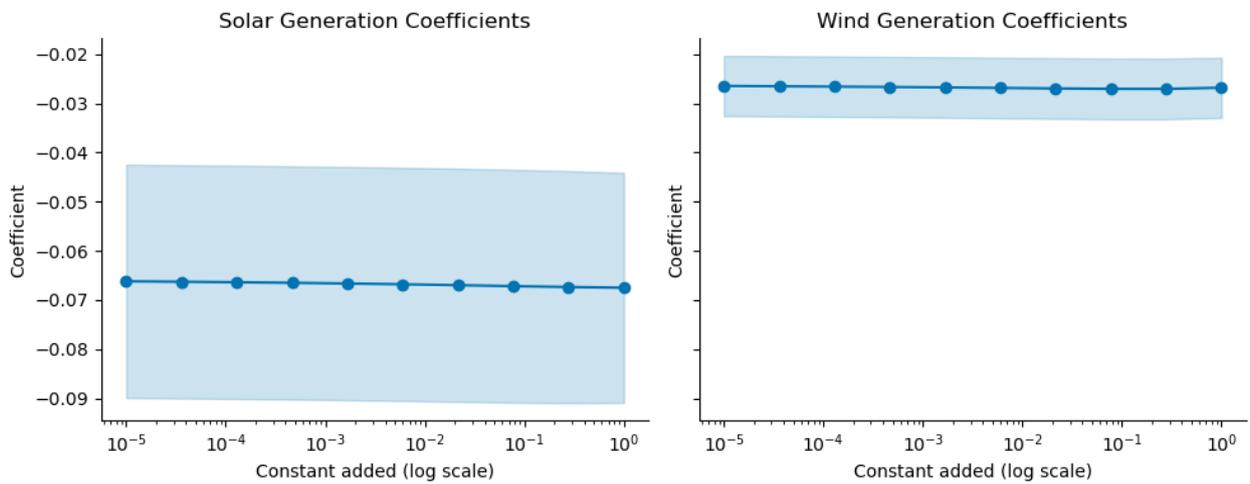

(b) Drop hours where power plant generation is 0.

Figure 7 | **Coefficients of solar and wind generation evaluated across a range of** $\Delta$ for CAISO. The x-axis is in log scale.

## C. Displacement effectiveness

The displacement effectiveness ($E$) is defined as:

$$E = \frac{\Delta P_{\text{observed}}}{\Delta P_{\text{expected}}}$$

where: - $\Delta P_{\text{observed}}$ is the observed reduction in emissions, - $\Delta P_{\text{expected}}$ is the expected reduction under perfect 1:1 displacement of thermal generation by renewables.

**Expected Emissions Reduction** Under ideal conditions, renewable generation displaces thermal generation directly, such that:

$$\Delta P_{\text{expected}} = G_r \cdot \text{EI}_{\text{thermal}}$$

where: - $G_r$ is the renewable generation, - $\text{EI}_{\text{thermal}}$ is the emissions intensity of the thermal plants being displaced.

**Observed Emissions Reduction** In practice, the observed reduction can exceed the expected reduction due to secondary effects, such as:

1. Avoidance of Start-Up and Shutdown Cycles: Renewable generation may stabilize the grid, reducing inefficient start-up and shutdown operations in thermal plants.
2. Improved Load Distribution: Renewables can enable thermal plants to operate closer to their optimal efficiency points, reducing emissions intensity per unit of output.
3. Displacement of Higher-Intensity Plants: In some regions, renewables displace thermal units with above-average emissions intensity, leading to greater-than-expected reductions.

Thus, the observed reduction can be expressed as:

$$\Delta P_{\text{observed}} = G_r \cdot \text{EI}_{\text{thermal}} + F$$

where $F$ represents the secondary effects described above.

**Displacement Effectiveness Greater than 1** When $F > 0$, the displacement effectiveness exceeds 1:

$$E = \frac{G_r \cdot \text{EI}_{\text{thermal}} + F}{G_r \cdot \text{EI}_{\text{thermal}}} = 1 + \frac{F}{G_r \cdot \text{EI}_{\text{thermal}}}$$

This condition occurs when the secondary effects ($F$) amplify the emissions reductions beyond what would be achieved under a simple 1:1 displacement.

**Implications** Displacement effectiveness greater than 1 highlights the importance of system-wide operational improvements and underscores the potential for renewables to drive emissions reductions through indirect mechanisms, such as improved plant dispatch efficiency and grid stability enhancements.